# Systolic Pressure in Different Percents of Stenosis at Major Arteries

Mohammad Reza Mirzaee, Omid Ghasemalizadeh, Bahar Firoozabadi, Meitham Dandaneband

**Abstract -** Modeling Human cardiovascular system is always an important issue. One of the most effective methods is using lumped model to reach to a complete model of human cardiovascular system. Such modeling with advanced considerations is used in this paper. Some of these considerations are as follow: Exact simulating of ventricles as pressure suppliers, peristaltic motion of descending arteries as additional suppliers, and dividing each vessel into more than one compartment to reach more accurate answers. Finally a circuit with more than 150 RLC segments and different elements is made. Then the verification of our complex circuit is done and at the end, obstruction as an important abnormality is investigated. For this aim different percents of obstruction in vital arteries are considered and the results are brought as different graphs at the end. According to physiological texts the citation of our simulation and its results are obvious. To earn productive information about arteries characteristics a 36-vessels model was chosen from biological sources.

## 1 Introduction

Diseases are always the major problem for human health. One of them is obstruction in different arteries which would cause abnormity in actions of different members. By making a strong model of whole human cardiovascular clinical studies would be done simpler. Such modeling would be achievable by using different method. One of the most effective ways to create a citable and accurate model is using electrical analogy. Another methods such as one or multi-dimensional modeling and experimental methods are beneficiary too, but they aren't the subject of this paper. In this context electrical analogy (lumped method) is used

Mohammad Reza Mirzaee is student of Sharif University of Technolofy (corresponding author; phone: +98-21-6165684; fax: +98 21 6000021; e-mail: mirzaee83@yahoo.com).
Omid Ghasemalizadeh is student of Sharif University of Technolofy (e-mail: alizadeh.omid@gmail.com).
Bahar Firoozabadi is associated professor of Biofluid Group of Sharif University of Technolofy (e-mail: firoozabadi@sharif.edu).
Meitham Dandaneband is student of American University of Sharjah (e-mail: b00028780@aus.edu).

with the goal of providing better understanding and simulation of blood characteristics in the human cardiovascular system which will lead to accurate answers as a result of exact modeling. Such modeling has been used in previous studies but not by such details that are used in our research. The first computer models describing the arterial system such as ascending aorta and carotids were introduced a multi-branched model of the arterial tree in a usable form for digital computers. By this method different physiological conditions became considerable. Later more detailed models were applied to reach more accurate results[1]. To analyze the human cardiovascular system mathematically, more simplified model should be considered to decrease the difficulty of investigating. A pulsatile-flow model of the left and right ventricles (as suppliers) and 2-segment aorta were constructed and the changes in flow behavior investigated. Later lumped (electrical analogy) model was developed to analyze cardiovascular systems easily with suitable accuracy. An electrical model which focused on the vessel properties was made by Young and his team[2]. One-dimensional axisymmetric Navier-Stokes equations for time dependent blood flow in a rigid vessel had been used to derive lumped models relating flow and pressure[3]. The effects of external factors on vessels wall were investigated in blood flow[4]. A complicated non-linear computer model for pressure changes and flow propagation in the human arterial system was drived[5]. The model had 55 arterial compartments and was based on one-dimensional flow equations to simulate effects of hydrodynamic parameters on blood flow. A computer model of stenosis was introduced and analyzed with a related software[6]. Later a simple model of lumped method was presented for human body[7].

This paper describes modeling of the whole human cardiovascular system using an extensive equivalent electronic circuit (lumped method). The main assumptions are taken from previous researches. But to reach more powerful and precise model, in this paper we have taken a quite different way to model the cardiovascular system and we had intention to develop our electrical models by applying more details from main arteries which more accurate results would be one of its effects. The model consists of about 150 RLC segments representing the arterial and cardiac systems that would be explained later. Using more compartments to complete arteries circuit would result more exact modeling. Respect to previous researches in electrical analogy, our modeling has more advantages and considerations. Also, body vessels equivalent compartments in the circuit is more detailed which would result crossing flow completely and without wasting through whole body. Furthermore, adding suppliers to model waveform movement of descending vessels makes our modeling more and more accurate and citable. Pressure graphs which are shown in results would confirm the above statements.

So investigating cardiovascular system faults and effects of diseases would accompany precise results.

Here is a 36-vessel body tree which is utilized in this research to model human cardiovascular system vessels. Also information about these vessels is shown in table 1.

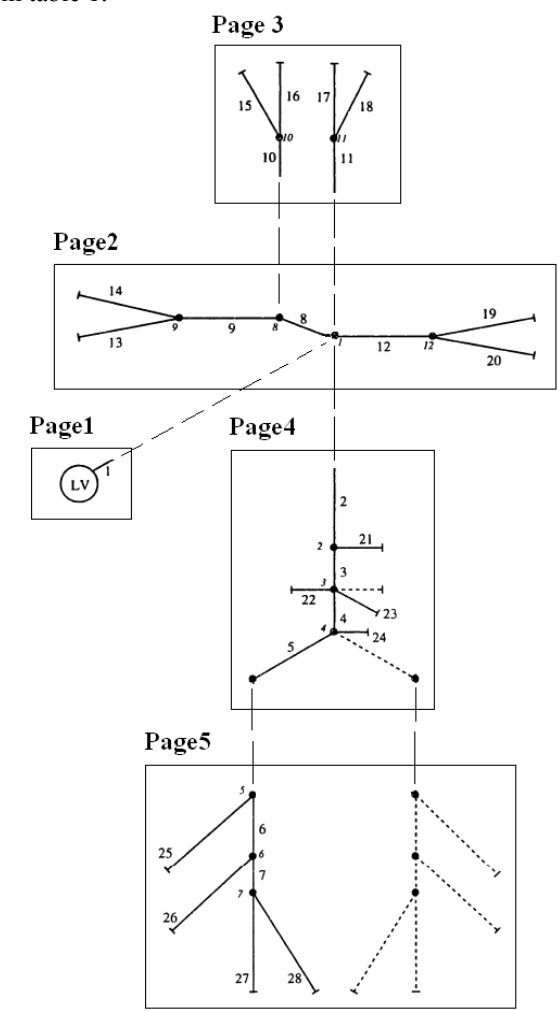

**Fig1 – 36-Vessels Body Tree**

## 2  Introducing Modeling Principles

To model human cardiovascular system we chose different equivalent electrical elements to express different mechanical properties of vessels, blood and heart.

In this modeling process atriums, ventricles, every blood vessel, set of all capillaries, arterioles and veins have been presented by some compartments consisting of a resistor, an inducer and a capacitor.

The number of compartments would be chosen by the purpose and the required accuracy .So more compartments would be used for main arteries.

Voltage, current, charge, resistance and capacitance, inductor in the electronic circuit are respectively equivalent to blood pressure, blood flow, volume, resistance, compliance and flow inertia in the cardiovascular system. Ground potential (reference for voltage measurements) is assumed to be zero as usual. The relation between mechanical properties of cardiovascular system and their equivalent electrical elements are as follow:

0.01ml/Pa = 1 µF (compliance - capacitance)
1 Pa.s$^2$/ml = 1 µH (inertia - inductor)
1 Pa.s/ml = 1 kΩ (resistance - resistance)
1mmHg = 1 volt (pressure - voltage)
133416 ml = 1A (volume - charge)

Following modeling is taken to introduce needed electrical elements for simulation[8].

Blood vessel resistance (R), depending on blood viscosity and vessel diameter, is simulated by resistors. This simulation has considered because blood viscosity will cause resistance against Blood flow crossing.

The blood inertia (L) is simulated by inductors. Reason of this consideration is variability of flow acceleration in pulsatile blood flow, so an inductor can model inertia of blood flow very clearly.

The vessel compliance (C) is considered For the reason of this simulation, it should be noted that by passing blood thorough vessels, the vessels would be expanded or contracted, so they can keep blood or release it, and this is exactly like what a capacitor does with electrical charges. By these statements each vessel is modeled by some compartments, which includes one resistance, one capacitor, and one inductor. The next step is to introduce a model to put these elements together. Below model is chosen to make the circuit[8].

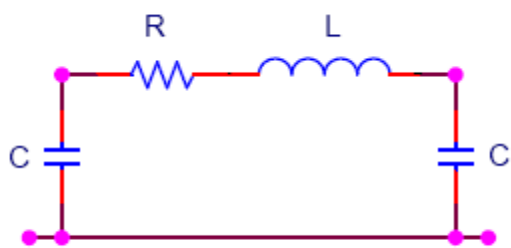

**Fig2 – Compartment shape and its elements**

Also, Quantities of Compartment's elements are easily achievable by using related equations in physiological texts which is available in references part[8]. Computed values of circuit elements are shown in table 1. Also, it shall be noted that substituting these quantities in their relevant elements should be done with adequate precision and in a special manner.

**Table 1 – Calculated Values for Elements of Circuit from artery parameters**

|    | Vessel Name | n | R (KΩ) | C (μF) | L (μH) |
|----|-------------|---|--------|--------|--------|
| 1  | Ascending Aorta | 27 | 0.000411 | 0.044883 | 0.007286 |
| 2  | Thoracic Aorta | 5 | 0.025155 | 0.430674 | 0.243014 |
| 3  | Abdominal Aorta | 2 | 0.033444 | 0.151642 | 0.213599 |
| 4  | Abdominal Aorta | 3 | 0.270026 | 0.087308 | 0.740451 |
| 5  | Common iliac | 6 | 0.390519 | 0.069687 | 0.890459 |
| 6  | Femoral Artery | 5 | 23.19873 | 0.014663 | 11.27735 |
| 7  | Anterior Tibial Artery | 1 | 46.97287 | 0.000277 | 6.686321 |
| 8  | Brachiocephalic | 1 | 0.144677 | 0.079312 | 0.469379 |
| 9  | R Brachial | 8 | 1.78196 | 0.060436 | 2.40721 |
| 10 | R Common Carotid | 6 | 0.973559 | 0.033019 | 1.315159 |
| 11 | L Common Carotid | 5 | 0.764939 | 0.025943 | 1.033339 |
| 12 | L Brachial | 1 | 12.72576 | 0.586882 | 18.9344 |
| 13 | R Radial | 5 | 63.76945 | 0.002583 | 13.70198 |
| 14 | R Ulnar | 5 | 20.8499 | 0.005343 | 7.783999 |
| 15 | R External Carotid | 5 | 5.174922 | 0.004234 | 2.724107 |
| 16 | R Internal Carotid | 5 | 3.778271 | 0.010145 | 2.871731 |
| 17 | L Internal Carotid | 5 | 3.778271 | 0.010145 | 2.871731 |
| 18 | L External Carotid | 5 | 5.174922 | 0.004234 | 2.724107 |
| 19 | L Radial | 5 | 21.12304 | 0.005413 | 7.885973 |
| 20 | L Ulnar | 5 | 62.94484 | 0.00255 | 13.5248 |
| 21 | Coeliac | 1 | 0.384992 | 0.010925 | 0.494247 |
| 22 | Renal | 1 | 5.284447 | 0.01138 | 3.008844 |
| 23 | Sup Mesenteric | 1 | 1.407178 | 0.07666 | 2.195783 |
| 24 | Inf Mesenteric | 1 | 61.85005 | 0.00501 | 13.28906 |
| 25 | Profundis | 1 | 38.62573 | 0.00921 | 17.22063 |
| 26 | Post Tibial | 1 | 258.7192 | 0.012262 | 70.875 |
| 27 | Ant Tibial | 1 | 2700.264 | 0.002839 | 224.8185 |
| 28 | Proneal | 1 | 980.1671 | 0.005771 | 139.5212 |

Where n is number of segments that the vessels are divided to.

Also we should note that heart with its moving muscles is the power supplier of blood circulation in whole body vessels.

To have a very exact modeling of blood and vessels behaviors, exact modeling of power suppliers is an important factor. So in our simulation left and right ventricles are modeled quite exactly the same as biological graph sources[9]. These are visible in verification parts.

Atriums are simulated as part of the venous system without any contraction. Atriums and ventricles can be modeled like vessels as a RLC segment. These two important parts of heart have two courses of action, one is resting position (diastole) which muscles would take their maximum volume and the other is acting position (systole) which muscles reach to their minimum size. Different heart shutters are modeled by using appropriate diodes, because shutters like diodes cross the flow in one direction. Considering this fact is important because in some parts of cardiovascular system, inverse current movement will cause to great danger to health. So to reach a good model, choosing appropriate diodes would be counted as an inseparable part.

Also, bifurcations are important cases to have accurate modeling. For simulating these parts of cardiovascular system a special method has been used[8]

In brief, it can be said that in bifurcations properties of jointed vessels will combine together in a complicated way to show the blood current division effects.

### 2.1 Peristaltic Motion of Vessels

When several simultaneous measurements are done at different points all along the aorta, it appears that the pressure wave changes shape as it travels down the aorta. Whereas, the systolic blood pressure actually increases with distance from the heart. Thus the amplitude of the pressure oscillation between systole and diastole, which is pulse pressure, nearly doubles. Thereafter, both PP (pulse pressure) and MAP (mean aorta pressure) decrease rapidly[4].

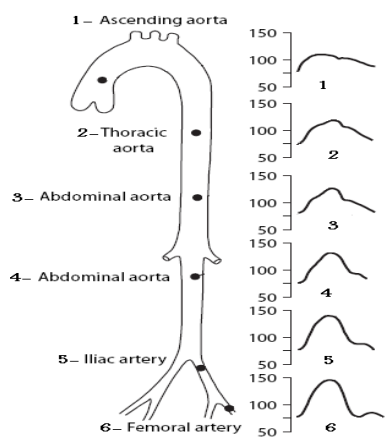

**Fig3 – The pressure changes thorough descending arteries caused by peristaltic motion**

Logically, wastes because of frictions and bifurcations should be caused decreasement of systolic pressure of descending arteries but as said above from biological texts, it is obvious from figure 3 [4], That by moving forward through these arteries the maximum point of pressure graph would be increased. The reason of this phenomenon is that these arteries have an additional pressure which is the result of waveform movement of vessels walls. These walls have a peristaltic motion which it would help the easier movement of blood in descending vessels.

To have more exact results from our circuit, in addition to precise simulation of left and right ventricles we used appropriate pressure suppliers to consider these peristaltic motions and their effects. To model these pressure suppliers in our electronic circuit for each artery and respect to its motion, appropriate voltage sources were chosen.

### 2.2 Obstruction Modeling

One of the most important goals of this research is probing different percents of constriction in main arteries. To have a good modeling of this phenomena real shape of arteries stenosis is brought in following pictures. It is obvious that obstruction in descending arteries such as thoracic, abdominal aorta, iliac, and femoral is so usual, so these vessels would be our main purpose.

As said above descending arteries are our main point of focus. So, exact pressure graph of them is so important. Because of this, considering peristaltic motion of these arteries is so urgent which it has been considered in the equivalent circuit and its explanations are implied in part 3.1. Simulated 3D model of obstruction for these arteries is shown in figure 4.

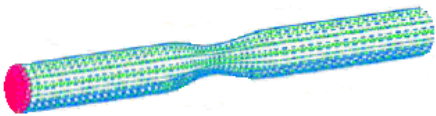

**Fig4 – 3D Shape of Obstruction in Main Arteries**

As it is obvious, obstruction would impose tighter path for blood flow to cross through the vessels. So the pressure downstream of the path would have lesser but upstream would provide more strong pressure to cross the flow.

## 3  Equivalent Circuit

This circuit includes five main parts that each one is applied for one part of body (The relation is obtainable from figure 1).

- Heart and ascending aorta circuit
- upper body circuit (hands and carotid)
- downer body circuit (descending arteries)

From circuit and its subdivisions it is vivid that each artery is divided into more than one RLC segments. In each part of circuit increasing the number of compartments would close our answers to real values because this work will increase the number of capacitors and consequently, their values would decrease, so this will lead to lesser leakage of current.

### 3.1  Heart and Ascending Aorta Circuit

- Ascending aorta and heart (ventricles, atriums, pulmonaries, shutters, and aorta artery).

Ascending aorta would be subdivided into 27 segments which elements quantities are shown in table 1.

Also an exact model of ventricles pressure has been used as the supply of power. The simulated and exact pressure graphs have been compared in Fig5 for right ventricle and Fig6 for left ventricle.

From the figures it is obvious that left and right ventricles pressure will change, in turn between 120-11 volt (mmHg) and 29-7 volt (mmHg).

### 3.2  Arteries Circuit except Ascending Aorta

- upper part of body (vessels upper than ascending aorta)
- downer part of body (vessels downer than ascending aorta)

The number of segments of other arteries would be chosen by their importance.

In these cases elements could be determined by using equation 1, 2 and 3. Also the calculated quantities are brought in table 1.

As said before voltage suppliers would be considered to model motion of vessels muscles. This should just be added to important descending arteries from aorta such as thoracic, abdominal aorta, iliac and femoral arteries.

For obstruction it should be noted, because each segment in obstruction region has its own radius, its elements quantities would be different.

### 3.3  Obstruction circuit

To model obstruction in main arteries different percents of constriction are chosen to obtain the results. It is obvious for each percent, different R, L, and C would be computed. Because of the multiplicity of obstruction percents and number of main arteries which are considered we avoid of bringing the whole data about our research but we will show the results in part 6.

To model this phenomenon we should change some RLC elements in specified vessels.

### 3.4  Additional Points

It should be noted that capillaries are so small but have an important role in cardiovascular system which without them the circulation of blood would not be completed. In our circuit marker "Vcc3" plays role of capillaries which connect arteries to veins. If these parts of circuit don't put correctly in their places, the modeling and its equivalent circuit wouldn't accompany real and exact answers.

It should be added that "VCC(x)"s, allude to co potential points.

It should be noted that by using more than 80 segments to model whole arteries there is no leakage of current in the system and arteries pressure graph are quite acceptable comparing to the input pressure. It means blood flow after exiting from left ventricle and crossing from arteries and without decreasing of its quantity, would enter veins from capillaries. By power of right ventricle it will pump to pulmonary and will pour to left ventricle again so the cycle would be completed.

Also model is capable of showing blood and vessels properties in different points. For example the pressure (voltage) and volume (charge) graphs can be obtained from the different points of the circuit easily. But in this circuit we would quietly focus on pressure graphs that are productive in clinical researches.

## 4  Verification of results

This section would be consisting of two parts.
- Confirmation of right and left ventricles pressure simulation as main suppliers.
- Confirmation of descending arteries pressure graph which is the result of exact modeling of peristaltic motion.

### 4.1 Ventricles verification

Fig5 shows the simulated pressure graph of right ventricle varying between 29–7 mmHg (volt). The real graph of right ventricle confirms our simulation results which are in complete accordance with physiological data of reference[9].

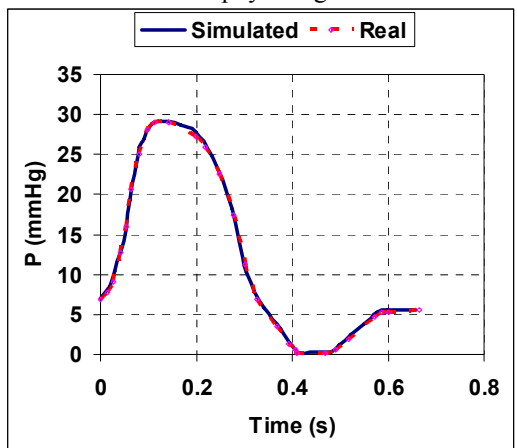

**Fig5 – simulated (solid line) and real (doted line) pressure for right ventricle**

The simulated pressure-time graph of left ventricle is shown in Fig6, where the waveform varies between 120–11 mmHg (volt). The results are in complete agreement with experimental observation of physiological text[9] which is brought in the same figure. The waveform starts from 11 mmHg and the peak is in 120 mmHg.

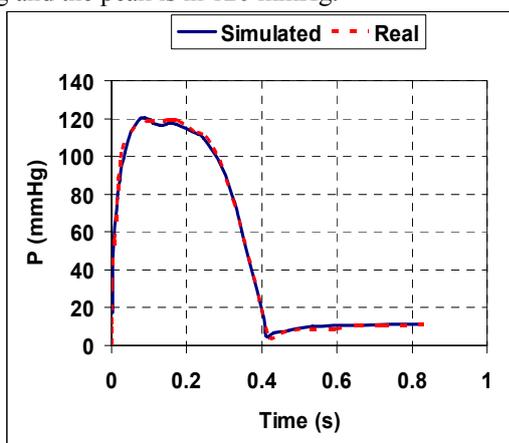

**Fig6 – simulated (solid line) and real (doted line) pressure for left ventricle**

### 4.2 Verification of peristaltic motion

The great work is done in this paper is reaching to exact pressure graphs of descending arteries that are obtainable below. This work is done by simulating the peristaltic motion of vessels wall. As result the pressure graph of arteries which have this motion such as thoracic, abdominal aorta, iliac, and femoral are shown in this part.

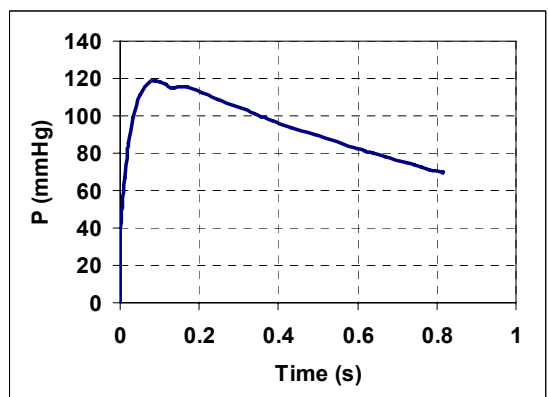

**Fig7 – calculated pressure for ascending aorta (1) artery**

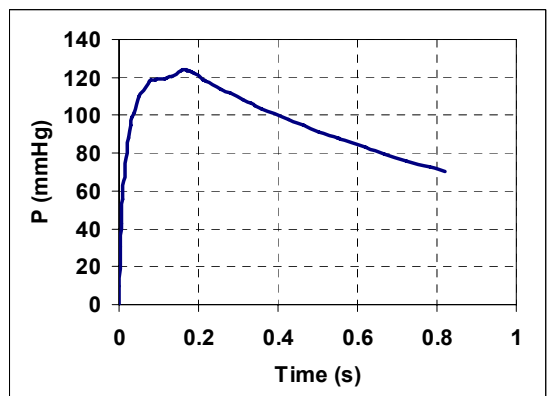

**Fig8 – calculated pressure for thoracic (2) artery**

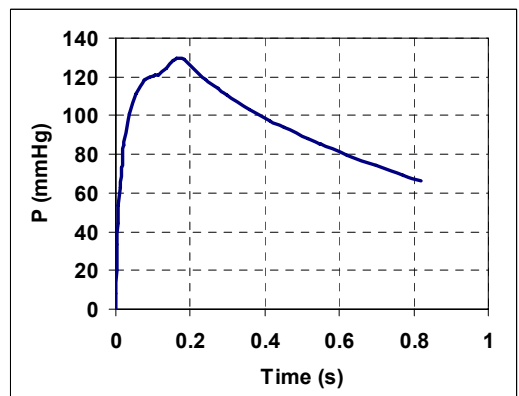

**Fig9 – calculated pressure for abdominal aorta (3) artery**

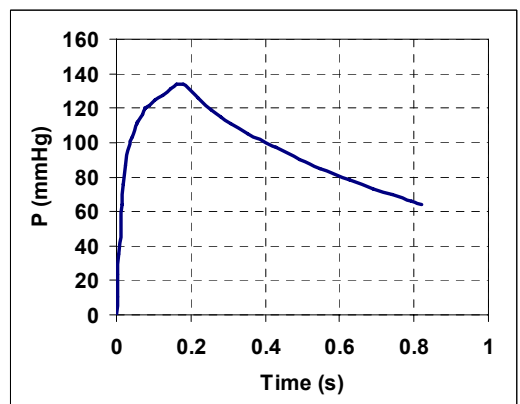

**Fig10 – calculated pressure for abdominal aorta (4) artery**

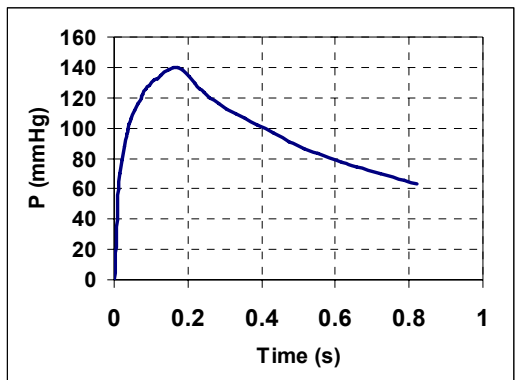

**Fig11 – calculated pressure for iliac (5) artery**

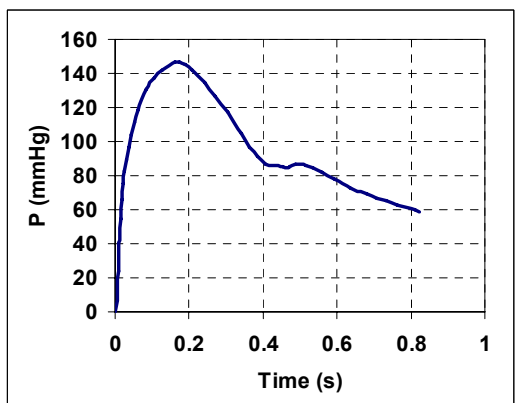

**Fig12 – calculated pressure for femoral (6) artery**

By referring to figures 7, 8, 9, 10, 11, 12, and comparing them with figure 3, it is easily notable that the calculated pressure quantities from our modeling is in great accordance with the real one. So by these results precision of our modeling would be greatly confirmed. For example pressure graph of aorta would be explained.

The calculated pressure changes of ascending aorta artery with 27 compartments are shown in Fig13. This graph shows that aorta pressure varies between 120–68 mmHg (volt) (systole-diastole) and the results are in exact agreement with physiological article[9] (Fig14). Even, two peaks in pressure graph of ascending aorta are earned the same as real measurements.

Because of these agreements between real and calculated graphs for ascending aorta, right and left ventricles the citation of our circuit and modeling would be verified.

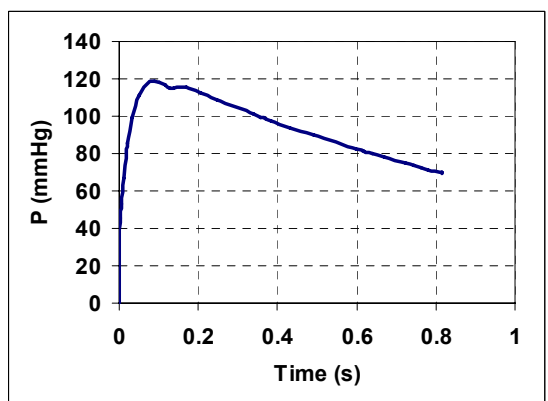

**Fig13 – Calculated Pressure for Ascending Aorta**

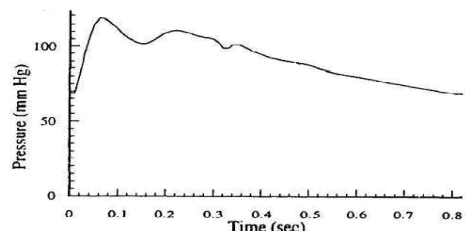

**Fig14 – Real Pressure for Ascending Aorta**

## 5   Results

After modeling peristaltic motion of descending arteries exactly and proving the citation of the vessels equivalent circuit in the verification part, obstruction with different percents in descending arteries such as thoracic, abdominal aorta, iliac, and femoral is investigated and the results for peaks (SBP) of their pressure graph are shown as follow.

The percents of obstruction are chosen from 30% to 80% except right femoral artery that its variation is taken from 10% to 70%. Reason of this decision is sensitivity of femora to stenosis even in low percents of obstruction, but other arteries don't show this in their behavior.

As said above the number of segments for each artery is distinct. To consider obstruction one segment for each named artery is chosen. This information are obtainable in table 2.

**Table 2 – Considered Length for Stenosis Region**

| Vessel Name | Length of Stenosis (Cm) |
|---|---|
| Thoracic | 3.7 |
| Upper Abdominal Aorta | 2.15 |
| Downer Abominal Aorta | 3.2 |
| Iliac | 3.2 |
| Right Femoral | 8.64 |

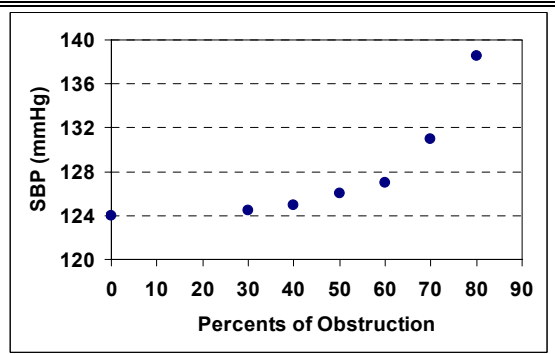

**Fig15 – Calculated SBP for Different Percents of obstruction in Thoracic**

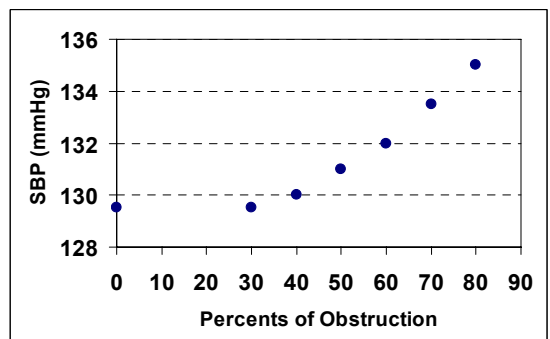

**Fig16 – Calculated SBP for Different Percents of obstruction in Upper Part of Abdominal Aorta**

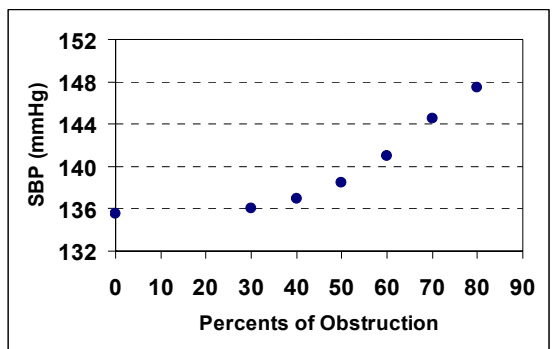

**Fig17 – Calculated SBP for Different Percents of obstruction in Downer Part of Abdominal Aorta**

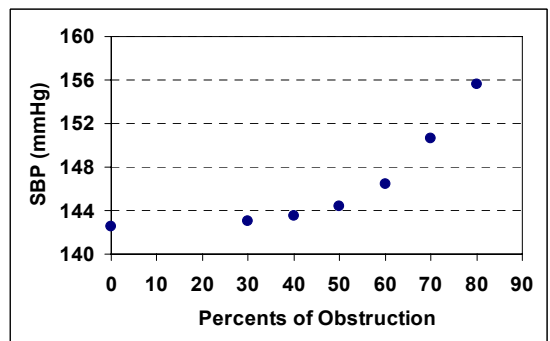

**Fig18 – Calculated SBP for Different Percents of obstruction in Iliac**

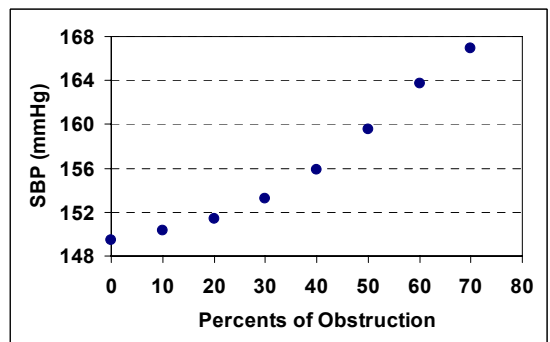

**Fig19 – Calculated SBP for Different Percents of obstruction in Right Femoral**

## 6  Conclusion

As it was shown in results confirmation the citation of our circuit is quite acceptable.

So it should be noted that using this complex electronic circuit to model human cardiovascular system with its all details, is so useful for studying of blood, different vessels and heart behaviors respect to each other and in different conditions such as health, diseases and abnormities. These abnormities may be obstructions, heart problems, vessels diseases or lots of different other things which are out of scope of this paper.

Finally it should be said that our circuit has this tendency to be more accurate and useful by adding more compartments and details to it.